\documentclass[%
showpacs,
 amsmath,amssymb,
prc,
twocolumn,
]{revtex4-1}
\usepackage{graphicx}
\usepackage{dcolumn}
\usepackage{bm}
\usepackage{hyperref}
\usepackage{xcolor}

\begin{document}
\preprint{APS/123-QED}

\title{Probing transfer to unbound states of the ejectile with weakly bound $^7$Li on $^{93}$Nb  }
\author{\large S.~K.~Pandit$^{1,2}$}
\email{sanat@barc.gov.in}
\author{\large A.~Shrivastava$^{1,2}$}
\author{\large K.~Mahata$^{1,2}$}
\author{\large N. Keeley$^{3}$}
\author{\large V.~V.~Parkar$^{1}$}
\author{\large P.~C.~Rout$^{1,2}$}
\author{\large K.~Ramachandran$^1$}
\author{\large I.~Martel$^4$}
\author{\large C.~S.~Palshetkar$^1$}
\author{\large A.~Kumar$^1$} 
\author{\large A.~Chatterjee$^1$} 
\author{\large S.~Kailas$^1$}
\affiliation{$^1$Nuclear Physics Division, Bhabha Atomic Research Centre, Mumbai - 400085, India}
\affiliation{$^2$Homi Bhabha National Institute, Anushaktinagar, Mumbai 400094, India}
\affiliation{$^3$National~Centre~for~Nuclear~Research,~ul.~Andrzeja~So\l tana~7,~05-400~Otwock,~Poland}
\affiliation{$^4$Departamento de F\'{i}sica Aplicada, Universidad de Huelva, E-21071 Huelva, Spain}

\date{\today}

\begin{abstract}
The  two-step process of transfer followed by breakup is explored by measuring a rather complete set of  exclusive data for reaction channels populating states in the ejectile continua of the $^7$Li+$^{93}$Nb system at energies close to the Coulomb barrier. The cross sections for $\alpha+\alpha$ events from one proton pickup were  found to be  smaller than those for $\alpha+d$ events from one neutron stripping and $\alpha+t$ events from direct breakup of $^7$Li. Coupled channels Born approximation and continuum discretized coupled channels calculations describe the data well and support the conclusion that the $\alpha+d$ and $\alpha+\alpha$ events are produced by direct transfer to unbound states of the ejectile.

\end{abstract}

\maketitle



Exploring  the properties of weakly-bound stable/unstable nuclei via transfer reactions is a topic of current interest~\cite{Keel09,Cant15} and also a focus of the next generation of high-intensity isotope-separator on-line (ISOL) radioactive ion beam facilities. Due to the low breakup threshold of such nuclei, population of the continuum is probable and consequently a large coupling effect is expected at energies around the Coulomb barrier. This may take place directly through inelastic excitation of the projectile (prompt or resonant breakup) or by nucleon transfer leaving the ejectile in an unbound state (transfer-breakup)~\cite{Keel09,Cant15,Queb74,Sign03,Shri06,Pako06,Sant09,Souz09,Luon11,Simp16,Rafi10,Luon13,Heim14,Hu16,Capu16}. The large positive Q-values for the transfer of neutrons from light neutron-rich projectiles to heavy targets also emphasize the role of neutron evaporation following transfer~\cite{chat08,Deyo05}. Exclusive measurements are essential to disentangle these reaction channels. Also, complete measurements of different reaction channels as well as theoretical calculations are required to understand the interplay between them.   Among the limited exclusive measurements aimed at studying  different breakup processes, very few data on absolute cross sections are available for direct breakup~\cite{Sign03,Shri06,Pako06,Sant09,Souz09,Heim14}, while for transfer-breakup absolute differential cross sections are only available for the neutron transfer channels at energies close to the Coulomb barrier~\cite{Shri06,Sant09}. 

Among the processes discussed above, investigation of the two-step reaction mechanism, viz., one nucleon transfer followed by breakup, is of current interest for the weakly-bound stable nuclei $^{6,7}$Li and $^9$Be~\cite{Shri06,Rafi10,Luon11,Simp16,Luon13,Otom13,Hu16}. This complex process needs the simultaneous understanding of both the breakup and transfer reactions. In an earlier measurement of the $^7$Li+$^{65}$Cu system~\cite{Shri06} it was observed that 1n-stripping leading to $^6$Li in its unbound 3$_1^+$ excited state is more probable than inelastic excitation of $^7$Li to its resonant states. In  recent measurements with $^7$Li~\cite{Luon11,Simp16}, the importance of 1p-pickup over the direct breakup of the projectile was highlighted while explaining the suppression of fusion at energies above the Coulomb barrier. Hence, understanding the mechanism of projectile breakup---whether direct or transfer breakup---is crucial while studying the reaction dynamics of weakly bound nuclei.

This communication reports the first simultaneous measurement of absolute differential cross sections for both  1p-pickup and 1n-stripping followed by  breakup of the ejectile as well as direct breakup of the weakly bound projectile over a wide angular range.  A simulation code has been developed using the Monte Carlo technique to interpret the observables of different  breakup processes and estimate the efficiency  for coincident detection of the breakup fragments. Angular distributions for elastic scattering and nucleon transfer to bound states have also been measured. Coupled channels Born approximation (CCBA) and continuum discretized coupled channels (CDCC) calculations which explain the large number of observables are presented.

The experiment was carried out at the Pelletron-Linac facility, Mumbai, with $^7$Li beams of 24, 28 and 30 MeV. A self-supporting $^{93}$Nb foil of thickness $\sim 1.75$ mg/cm$^2$ was used as a target. The requirements of high granularity to detect low-lying resonant states  and  large solid angle to measure low cross section events  were achieved using segmented large area Si-telescopes of active area $5 \times 5$ cm$^2$.  The $\Delta E$ detector (50 $\mu$m thick) was single-sided and the $E$ detector (1.5 mm thick) was double-sided with 16 strips allowing a maximum of 256 pixels. Two such telescopes, set 30$^\circ$ apart, were mounted at a distance of 16 cm from the target on a movable arm in a scattering chamber. In this geometry, the cone angle between the two detected fragments ranged from $1^\circ$ to $24^\circ$. The angular range $30^\circ$-$130^\circ$ (around the grazing angle) was covered by measurements at different angle settings. Three Si surface-barrier detector telescopes (thicknesses: $\Delta E$ $\sim$ 20-50 $\mu$m, $E$ $\sim$ 450-1000 $\mu$m) were used to obtain the elastic scattering angular distribution at forward angles ($25^\circ$-$40^\circ$) where the count rate is too high for the strip detectors to cope with. Two Si surface-barrier detectors (thickness $\sim 300$ $\mu$m) were kept at $\pm20^\circ$ for absolute normalization.  The detectors were calibrated using the known $\alpha$ energies from a $^{239}$Pu-$^{241}$Am-source and the $^7$Li~+~$^{12}$C reaction at 24 MeV~\cite{Park07}.

\begin{figure}
{\includegraphics[width=80mm]{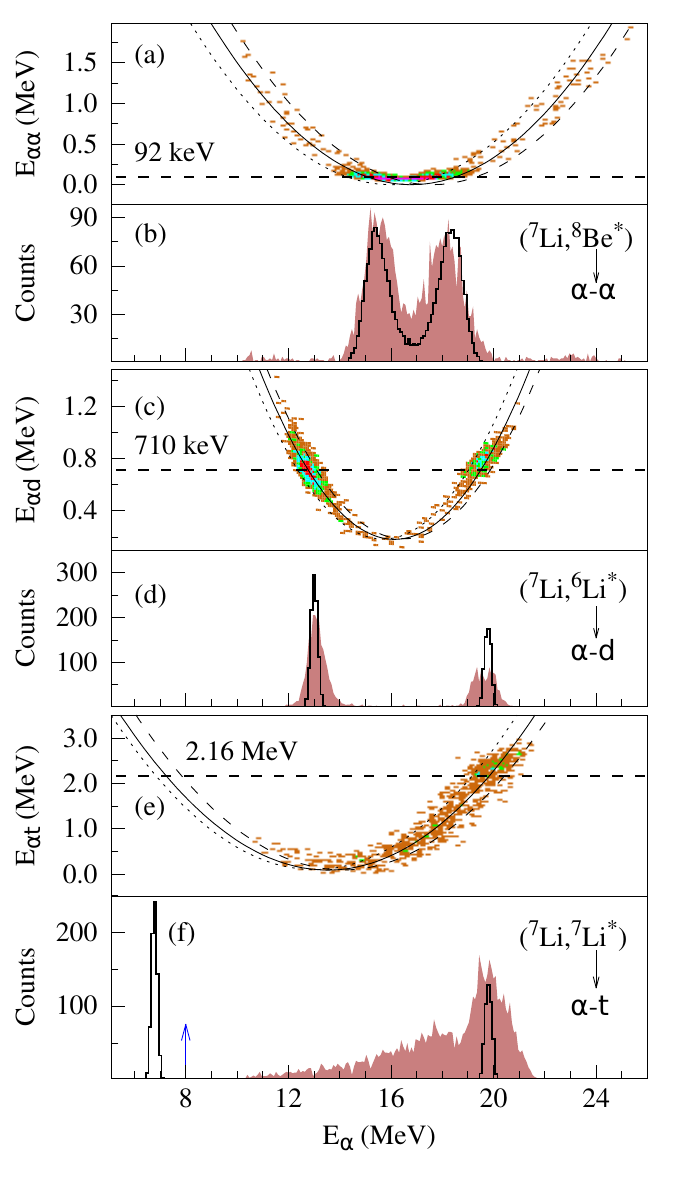}}
\vskip -3mm
\caption{\label{fig1}(Color online) Measured energy correlation spectra of breakup fragments for $^7$Li on $^{93}$Nb at $E_{\rm beam} = 28$ MeV and $\theta_{\rm lab}$~=~60$^\circ$. (a), (c) and (e): $E_{\alpha}$ vs.\ the relative energy $E_{\alpha\alpha}$, $E_{{\alpha}\textit{d}}$, and $E_{{\alpha}\textit{t}}$, corresponding  to  $\theta_{\rm rel}^{\alpha \alpha}$~=~$3^\circ$,   $\theta_{\rm rel}^{\alpha d}$~=~$10^\circ$ and  $\theta_{\rm rel}^{\alpha t}$~=~$15^\circ$, respectively. The shaded distributions in (b), (d) and (f) correspond to projections of the $\alpha$-particle energy for the data in (a), (c) and (e), respectively. The arrow on the x-axis indicates the detection threshold.   The  kinematical curves plotted as  dashed, solid and dotted lines  correspond to (a)  $E^{\star}$($^{92}$Zr) = 1.0, 3.0, and 5.0 MeV  (c)  $E^{\star}$($^{94}$Nb) = 0.0, 0.5 and 1.0 MeV and (e)  $E^{\star}$($^{93}$Nb) = 0.0, 1.0 and 2.0 MeV. $E_{\alpha}$ resulting from Monte Carlo simulations are shown as solid lines in (b), (d) and (f) for  fragments originating from $^8$Be(g.s.), $^6$Li(3$^+$) and $^7$Li(7/2$^-$), respectively  (see text for details).}
\vskip -4mm
\end{figure}

  Detected particles were tagged by kinetic energy ($E$), identity (\textit{A, Z}) and scattering angle ($\theta$, $\phi$) with respect to the beam axis.  The relative angles ($\theta_{\rm rel}$) between the fragments were calculated from the measured scattering angles ($\theta_1$, $\phi_1$; $\theta_2$, $\phi_2$). The fragments' mass, kinetic energy ($E_1, E_2$) and  $\theta_{\rm rel}$  were used to calculate their relative energy ($E_{\rm rel}$). The energy of the $\alpha$ particle $E_{\alpha}$ vs.\ the relative energy $E_{\alpha\alpha}$, $E_{{\alpha}\textit{d}}$ and $E_{{\alpha}t}$ is shown in~Fig.\ \ref{fig1} (a), (c) and (e), respectively, for $E_{\rm beam} = 28$ MeV and the center of the detector at $\theta_{\rm lab}$~=~$60^\circ$. The corresponding projections of the $\alpha$-particle energy are denoted by the shaded areas  in~Fig.\ \ref{fig1} (b), (d) and (f). The excitation energy of the ejectile prior to breakup was obtained by adding the breakup threshold to the measured $E_{\rm rel}$. The $E_{\rm rel}$ spectra for  $\alpha$+$\alpha$, $\alpha$+\textit{d} and $\alpha$+\textit{t} exhibit peaks at 92~keV, 710 keV and 2.16 MeV that correspond to the breakup of $^8$Be (g.s.), $^6$Li (2.18 MeV, 3$_1^+$) and $^7$Li (4.63 MeV, $7/2^-$), respectively.  The two peaks at high and low energy in the $E_{\alpha}$ spectra in Fig.\ \ref{fig1} (b) and (d) are due to $\alpha$ particles moving in the forward and backward direction in the rest frame of the ejectile prior to breakup.  For $\alpha$+\textit{t} coincidence events from breakup of the $7/2^-$ resonance, as shown in Fig.\ \ref{fig1} (e), only the high energy $\alpha$ events could be detected. The low energy $\alpha$ particles were stopped in the $\Delta E$ ($\sim$50 $\mu$m) detectors.

The excitation energy of  the target-like  nuclei was determined using the missing energy technique. For the transfer reactions, this  was found to peak around the energy $E^* = Q_{\rm gg} - Q_{\rm opt}$, as expected from semi-classical theory~\cite{Brin72}. Here $Q_{\rm gg}$ and $Q_{\rm opt}$ are the ground state and optimum Q-values, respectively. For 1p-pickup $E^*(^{92}$Zr) peaks at $\sim$3~MeV and for 1n-stripping $E^*(^{94}$Nb) peaks at $\sim$0.5~MeV. The data presented in~Fig.\ \ref{fig1} (a) and (b); (c) and (d); (e) and (f) correspond to excitation energies $E^*(^{92}$Zr) up to 5  MeV, $E^*(^{94}$Nb) up to 1 MeV and $E^*(^{93}$Nb) up to 2  MeV, respectively.

   The efficiency for the detection of fragments in coincidence was estimated using the Monte Carlo technique taking into account the excitation of the target as well as the ejectile, the \textit{Q}-value of the reaction, the energy resolution and detection threshold. The efficiency depends on the velocity of the ejectile prior to breakup as well as the relative velocity of the fragments~\cite{Maso92}.  The scattering angle of the ejectile prior to breakup was assumed to be isotropic. The scattered energy of the ejectile was calculated  using  kinematics. The breakup fragment emission in the rest frame of the ejectile was also considered to be isotropic. The velocities of each fragment in the rest frame of the ejectile were calculated using energy and momentum conservation laws. These velocities were added to the velocity of the ejectile prior to breakup to get their velocities in the laboratory frame. It was  checked whether both fragments hit two different vertical and horizontal strips. Events satisfying this condition were considered as detectable events for estimation of the efficiency. The conversion of the energy and scattering angle from the laboratory frame to the c.m. frame of the target-projectile in event-by-event mode automatically takes care of the Jacobian of the transformation.

The simulated kinetic energy spectra of the $\alpha$ particles resulting from different breakup processes are shown in~Fig.\ \ref{fig1}. The relative energy distributions for $\alpha$-$\alpha$, $\alpha$-$d$ and $\alpha$-$t$ breakup were assumed to be Gaussian, centered at 92, 710 and 2160 keV, respectively.  While the yields corresponding to the high and low energy $\alpha$ particle are symmetric for $\alpha$-$\alpha$ breakup, they are found to be asymmetric for $\alpha$-$d$ breakup.   Such an asymmetry was also observed in Ref.~\cite{Sant09} and different cross sections for the high and low energy fragments  were reported. However, in the present work, the observed asymmetry has been reproduced by the simulation  and consequently  consistent cross sections for the high and low energy $\alpha$ particles were obtained.  The asymmetry is found to arise from kinematic focusing, which depends on the fragment mass asymmetry, relative energy, relative angle and the scattering angle of the ejectile.

\begin{figure}
\includegraphics[width=80mm]{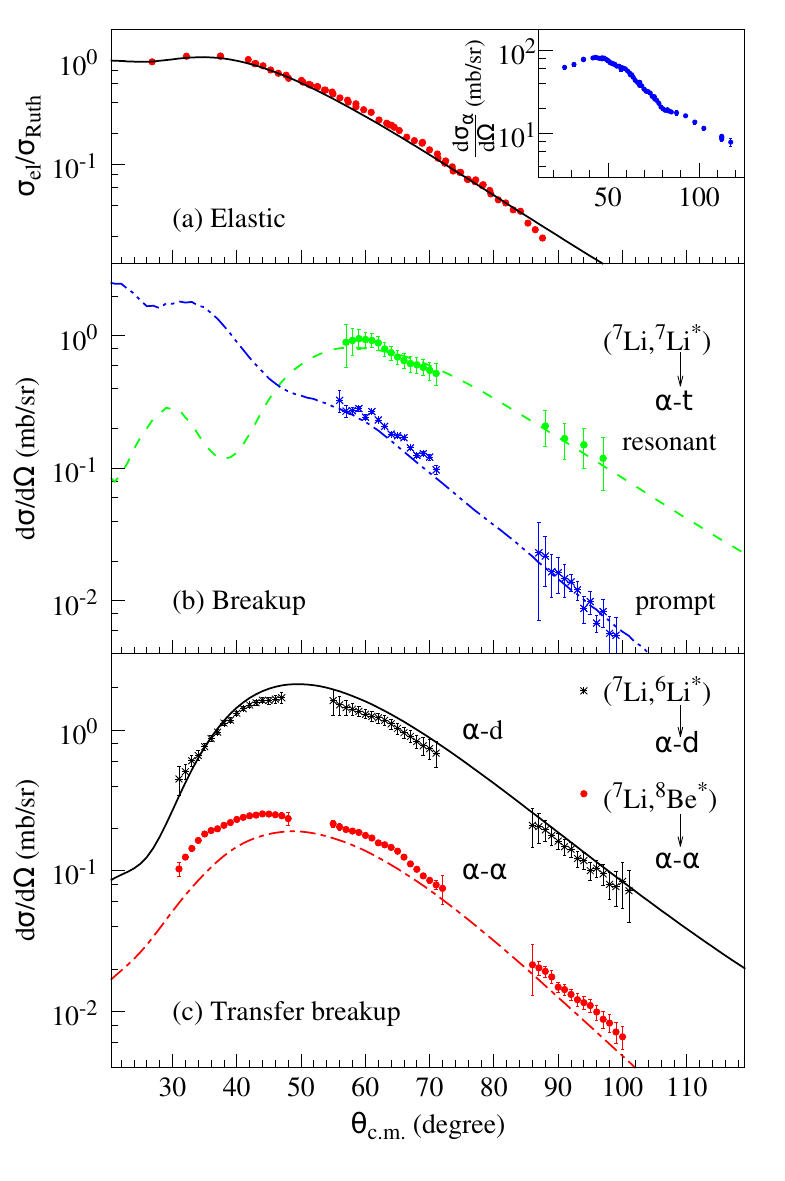}
\vskip -4mm
\caption{\label{fig2} (Color online) Measured inclusive and exclusive cross sections for the $^7$Li+$^{93}$Nb system at 28 MeV. (a) Elastic scattering data and the CDCC calculation. The inclusive cross section for $\alpha$ production is shown in the inset. (b) Prompt and resonant (from the 7/2$^-$ state) breakup of $^7$Li, shown as asterisks and filled circles, respectively. The CDCC results for prompt and resonant breakup are denoted by  dot-dot-dashed and dashed lines, respectively. (c) Exclusive data for  1p-pickup to $^8$Be($0_1^+$) and  1n-stripping to $^6$Li(3$_1^+$) are presented as filled circles and asterisks, respectively. The CCBA calculations for 1p-pickup and 1n-stripping are denoted by dot-dashed and solid lines, respectively.
}
\vskip -2mm
\end{figure}

The angular distributions of elastic scattering, projectile breakup and transfer followed by breakup for the $^7$Li+$^{93}$Nb system  at 28 MeV are shown in Fig.\ \ref{fig2}. The elastic scattering data are presented in~Fig.\ \ref{fig2} (a). The errors on the data points are due to statistics. The $^7$Li$^*$ $\rightarrow\alpha$+$\textit{t}$ breakup via the $7/2^-$ state and the continuum below this resonance are shown  in~Fig.\ \ref{fig2} (b). The cross sections for 1p-pickup leading to the $^8\mathrm{Be(g.s.)}$ are shown in~Fig.\ \ref{fig2} (c).  These data are restricted to $^{92}$Zr excitation energies up to 3.0 MeV,   as   information on the spectroscopic factors is available only in this energy range. For 1n-stripping, the cross sections for ${\alpha}$+$\textit{d}$ breakup events from the $^6$Li $3_1^+$ (2.18 MeV) state are shown in~Fig.\ \ref{fig2}~(c). Excited states of $^{94}$Nb up to 1.0 MeV were considered. The differential cross sections for $\alpha$+$d$ events from the breakup of $^6$Li formed after 1n-stripping are larger than those for $\alpha$+t events from the resonant breakup of $^7$Li, while those for $\alpha+\alpha$ events due to 1p-pickup forming $^8$Be are smaller. Integrated cross sections obtained assuming a Gaussian shape are listed in Table~\ref{tab:table1}. The total exclusive cross sections for $\alpha$ production form a small fraction of the inclusive cross section (inset Fig.\ \ref{fig2} (a)), indicating that the main $\alpha$ production mechanism is due to other processes, most likely fusion-evaporation and $t$-stripping/capture~\cite{Park07,Pako05,Shri06,Cant15}. Statistical model calculations using the code~{\sc PACE}~\cite{Gavr80} put the fusion-evaporation contribution to the total $\alpha$ yield at between 10\% ($E_\mathrm{beam} = 24$ MeV) to 20\% ($E_\mathrm{beam} = 30$ MeV).

\begin{figure}[h]
\includegraphics[width=85mm]{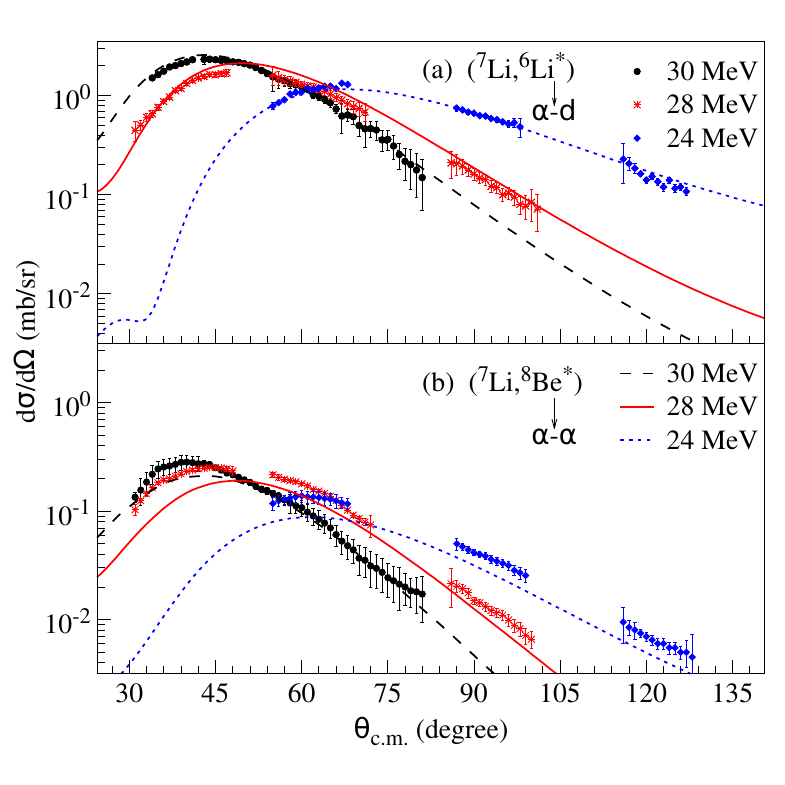} 
\vskip -4mm
\caption{\label{fig3} (Color online)   1n-stripping and  1p-pickup  cross sections for the $^7$Li+$^{93}$Nb system at 24, 28 and 30 MeV. (a) The $\alpha+d$ cross sections for the ($^7$Li,$^6$Li(3$_1^+$)) reaction. (b) The $\alpha+\alpha$ cross sections for the ($^7$Li,$^8$Be($0_1^+$)) reaction.  In both panels the dotted, solid and dashed lines correspond to the CCBA calculations at 24, 28 and 30 MeV, respectively.}
\end{figure}

Cross sections for 1n-stripping (Q$_{gg}$=$-20$ keV) populating the $^6$Li($3_1^+$) resonance and 1p-pickup (Q$_{gg}$=~11.21 MeV) populating $^8$Be(g.s.) at 24, 28 and 30 MeV are compared in~Fig.\ \ref{fig3} and Table~\ref{tab:table1}. The angular distributions for both reactions are bell shaped with peaks at the respective grazing angles for the different beam energies. The cross sections for 1p-pickup (with  a large positive Q$_{gg}$)  are found to be  smaller than those for 1n-stripping at all energies, which may
be attributed to poor kinematical matching~\cite{Brin72}. The $\alpha$+$t$ events from the $^7$Li(7/2$^-$) state could not be detected at 24 and 30 MeV due to the  detection threshold. Cross sections for 1n-stripping  populating $^6$Li in its ground state  and $^{94}$Nb (E$^*\le$1 MeV) were obtained independently from the inclusive data and are listed in Table~\ref{tab:table1}.


Two sets of calculations were performed to analyze the data. Calculations for elastic scattering and direct breakup were carried out within the CDCC formalism using the cluster folding model of $^7$Li.  The CCBA formalism was employed for transfer-breakup processes, using potentials that describe the elastic scattering data.  The code {\sc Fresco}~\cite{Thom88} was used in all cases.

The $^7$Li $\rightarrow$ $\alpha$+$t$ breakup data at 28 MeV were analyzed with CDCC calculations, similar to those described in Ref.\ \cite{Beck07} except that
the continuum bins were of width $\Delta k = 0.1$ fm$^{-1}$ with $k_\mathrm{max} = 0.8$ fm$^{-1}$ and $\alpha$+$t$ relative
angular momenta of $L=0$--$4$ and couplings up to multipolarity $\lambda = 4$ were included. The $\alpha$+$^{93}$Nb and $t$+$^{93}$Nb optical potentials required as input to the Watanabe-type folding potentials were taken from the global parameterizations of Refs.\ \cite{Avri94} and \cite{Becc71}, respectively. To obtain the best fit to the elastic scattering data the real and imaginary depths of
these potentials were renormalized by factors of 0.6 and 0.8, respectively. The results are compared to the elastic scattering and
breakup data in Fig.\ \ref{fig2} (a) and (b), respectively.

The $\alpha$+$d$ coincidence data were analyzed with CCBA calculations of the $^{93}$Nb($^7$Li,$^6$Li$^*$)$^{94}$Nb
1n-stripping reaction. In addition to the transfer couplings inelastic excitations of the $^7$Li($1/2^-$) and the $^6$Li($3_1^+$) excited states were included in the entrance and exit partitions, respectively. The entrance channel optical potentials were  based on the global $^7$Li parameters of Ref.\ \cite{Cook82} with real and imaginary depths readjusted to fit the elastic scattering data after the inclusion of the $^7$Li couplings. The $3/2^-$(g.s.) and $1/2^-$(0.478 MeV) states of $^7$Li were treated as members of a $K=1/2$ rotational band. The $B(E2;3/2^- \rightarrow 1/2^-)$ was taken from Ref.\ \cite{Well85} and the nuclear deformation length $\delta_2 = 2.4$ fm obtained by fitting the inelastic scattering data of Ref.\ \cite{Puig79}. The $^6$Li $B(E2; 1_1^+ \rightarrow 3_1^+)$ was taken from \cite{Ajze88} and the nuclear deformation length $\delta_2 = 1.9$ fm obtained by fitting the sequential breakup data of Ref.\ \cite{Vanv87}. The $^6$Li $1_1^+$ and $3_1^+$ states were assumed to be members of a $K=1$ rotational band, with the exception that reorientation of the $1_1^+$ ground state was omitted due to the very small quadrupole moment of this state. The exit channel optical potentials employed the $^6$Li global parameters of Ref.\ \cite{Cook82} with real and imaginary well depths adjusted to recover the same elastic scattering angular distributions at the appropriate energies when the $^6$Li excitations were included. Spectroscopic factors for the $\left<^7\mathrm{Li}|^6\mathrm{Li}+n\right>$  and $\left<^{94}\mathrm{Nb}|^{93}\mathrm{Nb}+n\right>$ overlaps were taken from Refs.\ \cite{Cohe67} and \cite{Moor69}, respectively.  Transfers between the $^7$Li $3/2^-$ ground state and the $^6$Li $1_1^+$ and $3_1^+$ states, the $^7$Li $1/2^-$ state and the $^6$Li $1_1^+$ state, plus a total of eight states in $^{94}$Nb (the most strongly populated according to Ref.\ \cite{Moor69}) were included. The results are compared to the data in Figs.\ \ref{fig2} (c) and \ref{fig3} (a).

Similar CCBA calculations were performed for the $\alpha$+$\alpha$ coincidence data considering the $^{93}$Nb($^7$Li,$^8$Be)$^{92}$Zr
1p-pickup process. Entrance channel potentials were as described above.
 Exit channel optical potentials used $^7$Li global parameters \cite{Cook82} since no $^8$Be potentials are available. Pickup to
the 0.0 MeV $0_1^+$ resonance of $^8$Be and the $0^+$ (g.s.), $5^-$ (2.45~MeV) and $4^-$ (2.74~MeV) states of $^{92}$Zr
were included. The spectroscopic factor for the $\left<^8\mathrm{Be}|^7\mathrm{Li}+p\right>$ overlap was taken from Ref.\
\cite{Cohe67} and those for the $\left<^{93}\mathrm{Nb}|^{92}\mathrm{Zr}+p\right>$ overlaps were obtained by fitting the
$^{93}$Nb($d$,$^3$He) data of Ref.\ \cite{Ches74}, yielding values of C$^2$S = 1.4, 1.0 and 1.0 for the $0^+$, $5^-$ and
$4^-$ states, respectively (N.B.\ the value of 10.8 in Ref.\ \cite{Ches74} for the $0^+$ state appears to be an error, since
it is inconsistent with the data plotted on their Fig.\ 8). The results are compared to the data in Figs.\ \ref{fig2} (c) and \ref{fig3} (b).

\begin{table}
\caption{\label{tab:table1}Cross sections for various channels in the $^7$Li+$^{93}$Nb system;
$\sigma_\mathrm{cal}$ denotes the results of CDCC ($^7$Li$^*$(7/2$^-$)$\rightarrow \alpha+t$) and CCBA (other reactions) calculations, see text.}
\begin{ruledtabular}
\begin{tabular}{lcc}
Channel&$\sigma_\mathrm{exp}$ (mb) &$\sigma_\mathrm{cal}$ (mb)\\
\hline
\textbf{E$_\mathrm{beam}$=24 MeV}\\
$\alpha$-inclusive & 273 $\pm$ 40 &  \\
$^8$Be(g.s.){\footnote{E$^*(^{94}$Nb)$\le$3 MeV}}$\rightarrow \alpha+\alpha$ & ~0.5 $\pm$ 0.1 & 0.36 \\
$^8$Be(g.s.){\footnote{E$^*(^{94}$Nb)$\le$5 MeV}}$\rightarrow \alpha+\alpha$ & ~1.0 $\pm$ 0.2 &   \\
$^6$Li$^*$(3$_1^+$)$\rightarrow \alpha+d$ & ~5.2 $\pm$ 0.5 & 5.5 \\
$^6$Li(g.s.) & 9.9 $\pm$ 1.0 & 9.8 \\
$\sigma$-reaction &  & 1121 \\

\textbf{E$_\mathrm{beam}$=28 MeV}\\
$\alpha$-inclusive & 321 $\pm$ 48 &  \\
$^8$Be(g.s.)$^\mathrm{a}$$\rightarrow \alpha+\alpha$ & ~0.7 $\pm$ 0.1 & 0.56 \\
$^8$Be(g.s.)$^\mathrm{b}$$\rightarrow \alpha+\alpha$ & ~1.3 $\pm$ 0.2 &   \\
$^6$Li$^*$(3$_1^+$)$\rightarrow \alpha+d$ & ~5.8 $\pm$ 0.4 & 6.2 \\
$^6$Li(g.s.) & 11.0 $\pm$ 1.2 & 10.9 \\
$^7$Li$^*$(7/2$^-$)$\rightarrow \alpha+t$ & ~3.3 $\pm$ 0.6 & 2.9 \\
$\sigma$-reaction &  & 1310 \\
\textbf{E$_\mathrm{beam}$=30 MeV}\\
$\alpha$-inclusive & 340 $\pm$ 52 &  \\
$^8$Be(g.s.)$^\mathrm{a}$$\rightarrow \alpha+\alpha$ & ~0.6 $\pm$ 0.1 & 0.53 \\
$^8$Be(g.s.)$^\mathrm{b}$$\rightarrow \alpha+\alpha$ & ~1.5 $\pm$ 0.2 &   \\
$^6$Li$^*$(3$_1^+$)$\rightarrow \alpha+d$ & ~6.2 $\pm$ 0.4 & 6.2 \\
$^6$Li(g.s.) & 11.2 $\pm$ 1.5 & 10.3 \\
$\sigma$-reaction &  & 1489 \\
\end{tabular}
\end{ruledtabular}

\end{table}

We now discuss the current investigation in the context of recent work in this area. A study of the quasi-elastic excitation function and barrier distribution of the $^7$Li+$^{144}$Sm system found the role played by neutron transfer followed by breakup in the reaction mechanism to be important as a two-step process~\cite{Otom13}.  An investigation of the 1p-pickup reaction using the classical dynamical model~\cite{Diaz07} has recently been reported for the $^7$Li+$^{58}$Ni system and further measurements were suggested fully to understand the influence of breakup processes---both direct and transfer induced---on fusion~\cite{Simp16}.  In the present work we have measured absolute cross sections for all major reaction channels leading to breakup processes for a single system, $^7$Li+$^{93}$Nb. The rather complete nature of the data set, including elastic scattering, combined with the calculations presented here, enables us to make the following conclusions about both the mechanism and the relative importance of these processes.  

The CCBA calculations agree well with the  $\alpha$+$d$  and $\alpha + \alpha$ coincidence data, both in terms of the shapes of the angular distributions and the absolute cross sections (see Table \ref{tab:table1} and Fig.\ \ref{fig3}). Population of the $^6$Li ground state is also  reproduced by the same calculations. Calculations omitting direct population of the $^6$Li ground state confirmed that transfer to the $^6$Li $1_1^+$ state followed by excitation of the $3_1^+$ resonance makes a negligible contribution, in agreement with Ref.\ \cite{Shri06}. Taken together, the data and calculations show unambiguously that the mechanism producing the $\alpha$+$d$ coincidences is direct 1n-stripping to the unbound $^6$Li $3_1^+$ state and the origin of the $\alpha + \alpha$ coincidences is 1p-pickup to $^8$Be(g.s.). As seen from Table \ref{tab:table1}, the cross sections of 1n-stripping  account  for $\sim 2$ \% and 1p-pickup  only $\sim 0.8$ \% (recall that each $^8$Be contributes two $\alpha$ particles to the total yield) of the inclusive $\alpha$ yields.

The $\alpha + t$ coincidence data are reproduced well by the CDCC calculations. The calculated total $^7$Li$^*$ $\rightarrow \alpha$+$t$ breakup cross section (16.8 mb) accounts for $\sim 5$ \% of the inclusive $\alpha$ yield  at E$_\mathrm{beam}$= 28 MeV, almost twice the combined contribution of 1n-stripping and 1p-pickup.

Overall, the combination of measurements and calculations suggests that at most only $\sim 8$ \% of the inclusive $\alpha$ yield can be accounted for by these processes. Thus, even allowing for population of the $^6$Li and $^8$Be non-resonant continua and the broad $^8$Be $2_1^+$ resonance the main $\alpha$-particle production mechanism must be due to other sources, most likely fusion-evaporation and $t$-stripping/capture~\cite{Park07,Pako05,Shri06,Cant15}.

In summary, the present work reports for the first time a detailed study of the various breakup mechanisms---1p-pickup and 1n-stripping to unbound states of the ejectile and direct breakup---for the same system at energies close to the Coulomb barrier. CDCC and CCBA calculations were performed to analyze a comprehensive data set comprising elastic scattering, direct breakup and transfer-breakup. These calculations confirmed that the $\alpha$+$d$  and $\alpha$+$\alpha$ coincidences mostly result from direct 1n-stripping to the $^6$Li $3_1^+$ state and 1p-pickup to the $^8$Be $0_1^+$, respectively. The present result also established that the main $\alpha$ production mechanism must be due to other processes, presumably fusion-evaporation and $t$-stripping/capture.  Many reactions with low energy unstable radioactive ion beams from newly available ISOL facilities are expected to be of similar nature, see e.g.\ a recent study of the $^7$Be+$^{58}$Ni system \cite{Maz15}.  The present rather complete data sets for the breakup and transfer-breakup mechanisms for a stable weakly bound nucleus plus the theoretical analysis provide an important benchmark in this respect.


We thank the Mumbai Pelletron-Linac accelerator staff for providing steady and uninterrupted beam and Mr. P. Patale for help during the experiment. The authors would like to thank Prof. K. W. Kemper for a careful reading of the manuscript and useful comments. 

\end{document}